\documentclass[doublecol,linenumbers]{epl2}
\title{Smart Polymeric Recognition of a Hexagonal
{Monolayer}}
\author{B. Liewehr\inst{1,2} \and M. Bachmann\inst{2}}
\shortauthor{B. Liewehr and M. Bachmann}
\institute{
\inst{1} Institute of Physics, University of Rostock,
Albert-Einstein-Stra{\ss}e 23, D-18059 Rostock, Germany\\
\inst{2} Soft Matter Systems Research Group, Center for Simulational
Physics, Department of Physics and Astronomy, University of Georgia, 
Athens, GA 30602, USA
}

\abstract{
We investigate the adsorption of a flexible polymer at a 
{hexagonally patterned monolayer}. All conformational polymer 
phases are identified, which enables the 
construction of a hyperphase diagram, parameterized by 
temperature and monolayer adsorption strength. The energy scale associated 
with 
the adsorption strength is a material parameter of the hybrid system that 
generically accommodates the behavior of entire classes of polymers 
interacting with hexagonal substrates. We also discuss 
a bridge-building mechanism for the formation of unique layered 
polymer 
structures with potential for applications in nanoscale transport. 
High-quality data sets necessary for the
statistical analysis of the structural 
phase behavior of the system were obtained in extensive generalized-ensemble 
Monte Carlo computer simulations.
}

\begin{document}
\nolinenumbers
\maketitle 
\section{Introduction} Despite their huge structural phase space, flexible 
polymers have the 
impressive capability of undergoing distinct structural transitions. Most 
studied is the coil-globule or $\Theta$ collapse transition that separates 
entropically dominant random, extended structures from the phase of globular 
conformations~\cite{khokhlov1,degennes1,grass1}. Globules, which are also 
long-range disordered, but compact and potentially ordered at short 
distances, are reminiscent of liquid droplets.

Much less understood is the crossover from these 
globules to ordered (such as crystalline or quasicrystalline icosahedral and 
decahedral) or disordered (amorphous) compact, solid phases. 
One reason is that features of 
this qualitative change in behavior are similar to first-order liquid-solid 
transitions. These are particularly difficult to study for finite systems 
because of significant surface effects. In meticulous studies, the 
relation between the structural 
characteristics, known from Lennard-Jones clusters of 
atoms~\cite{whales1,doye1,sbj1}, and the polymer chain length as well as 
temperature could be established for a coarse-grained model of finite
flexible polymers~\cite{sbj1,sslb1,mb1}.

In a different line of research, the adsorption properties of polymers at 
planar~\cite{vrbova1,prellberg1,ivanov1,mbj1} and 
nonplanar~\cite{milchev1,srebnik1,vb1} substrates have been of substantial 
interest, since hybrid 
organic-inorganic interfaces promise ``smart material'' properties 
with potential for nanosensory devices and other functional 
applications~\cite{hu1,su1}. 

The recent development of remarkable experimental techniques that made the 
exfoliation of atomic monolayers like graphene sheets possible created 
vibrant research fields in the materials 
sciences~\cite{geim1}. The detailed investigation of mechanical and 
electronic properties of individual, hexagonally structured atomic layers 
advanced the knowledge of properties of two-dimensional  
quasispherical and quasicylindrical atomic monolayer structures such as 
fullerenes and carbon nanotubes~\cite{dressel1}, respectively.

In this Letter, we combine these worlds and investigate the generic 
structural phases of a flexible, elastic polymer interacting with a 
{hexagonal layer attached to a homogeneous substrate. Such a 
monolayer can be created by depositing or covalently binding atoms or small 
molecules to the underlying substrate.} Flexible polymers can more easily 
adapt 
to their environment than other classes of polymers and thus enable the 
identification and characterization of the multitude of possible structural 
transitions in adsorption processes under varied conditions.
We study the phase changes by systematically
modifying the energy scale associated with the adsorption strength between 
monomers and vertices of the hexagonal layer under the 
influence of thermal fluctuations. Advanced replica-exchange 
simulations~\cite{sw1,huku1,geyer1} were performed and a variety of 
thermodynamic quantities and order parameters were analyzed statistically. 
Locations of extremal fluctuations of these quantities reveal qualitative 
changes in phase behavior and can be used for the localization and 
characterization as structural transitions. 

The multiple transition lines 
thus identified are used to construct the hyperphase diagram of the 
system projected into the combined spaces of temperature and adsorption 
strength. The latter is an effective material parameter. This result is  
crucial for any future design of hybrid materials composed of soft polymer 
components and solid substrates with hexagonally shaped surface 
layer. As a consequence of the increasingly more specific adaptation 
process to the hexagonal surface structure at larger adhesion strengths, 
novel polymer phases occur. Intermolecular 
forces require the formation of local bridges between 
monomers located in the center of the hexagons under the constraint of 
limited elasticity of the bonds connecting adjacent monomers. The formation 
of characteristic highly oriented surface patterns of polymer conformations 
adsorbed at such substrates have not yet been investigated in much detail, 
although these substrate-induced stable polymer structures offer potential for 
nanoscale transport applications.

\section{Modeling and Simulation} 
\begin{figure}
\centerline{\includegraphics[width=7cm]{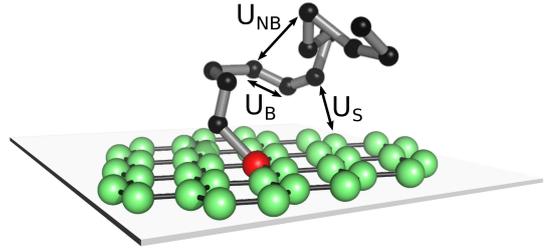}}
\caption{Model of a flexible polymer with one end monomer (red) grafted onto 
a rigid 
 {hexagonal monolayer on} an impenetrable substrate. The 
relevant potential 
energies for pairs of monomers and of monomers interacting with the 
 {attractive sites of the monolayer} 
are also symbolized.  {Note that connections between monolayer sites 
are only shown as guides to the eyes to visually enhance its hexagonal 
structure and do not represent ``bonds''.}}
\label{fig:model}%
\end{figure}

In our generic model of the hybrid system, 
the flexible polymer is grafted 
with one end at a {hexagonally patterned monolayer of attractive 
sites on an otherwise impenetrable substrate} (see Fig.~\ref{fig:model}). We 
employ a 
recently introduced model for flexible polymers~\cite{qlkpwb1}, in which the 
energy of a conformation, represented by the coordinate vector 
$\mathbf{X}=(\mathbf{r}_1, \mathbf{r}_2,\ldots,\mathbf{r}_N)$ for a linear 
chain with $N$ monomers, is given by the contributions from nonbonded (NB) 
and bonded (B) pairs of monomers, respectively, 
\begin{equation}
E_\mathrm{poly}(\mathbf{X})=\sum_{i<j} 
U_\mathrm{NB}(r_{ij})+\sum_i
U_\mathrm{B}(r_{i\, i+1}),
\end{equation}
where $r_{ij}=|\mathbf{r}_i-\mathbf{r}_j|$ is the distance between monomers 
$i$ and $j$. Denoting the standard Lennard-Jones (LJ) potential with energy 
scale $\epsilon$ and van der Waals radius $\sigma$ by 
\begin{equation}
U_\mathrm{LJ}(r)=4\epsilon[(\sigma/r)^{12}-(\sigma/r)^6],
\end{equation}
pairs of nonbonded monomers interact via the potential 
\begin{equation}
U_\mathrm{NB}(r)=(U_\mathrm{LJ}(r)-U_\mathrm{shift}) 
\Theta(r_\mathrm{c}-r),
\end{equation}
where $\Theta(r)$ is the Heaviside step function. 
The cut-off distance is $r_\mathrm{c}=2.5\sigma$ and 
$U_\mathrm{shift}=U_\mathrm{LJ}(r_\mathrm{c})$ avoids a discontinuity of the 
potential at $r=r_\mathrm{c}$. The minimum value of the potential is located 
at $r_0=2^{1/6}\sigma$. For bonded monomers, we choose a combination 
of the FENE (finitely extensible nonlinear elastic)~\cite{FENE1,FENE2,FENE3} 
and the LJ potential in the following form: 
\begin{eqnarray}
&&\hspace*{-10mm}U_\mathrm{B}(r)=-\epsilon 
KR^2\ln[1-(r-r_0)^2/R^2]/2+\eta[U_\mathrm{LJ}
(r)+\epsilon]\nonumber \\ 
&&\hspace*{4mm}-(U_\mathrm{shift}+\epsilon).
\end{eqnarray}
The parameter $\eta$ controls the symmetry and width of the 
potential~\cite{qlkpwb1}. In this study, we choose $\eta=1/10$. The other 
dimensionless parameters are set to $K=98/5$ and $R=3/7$~\cite{qlkpwb1}. The 
minimum-potential distance, chosen to be $r_0\equiv 1$ in the simulations, 
sets 
the basic length scale.  {It represents the energetically optimal 
distance between the centers of two monomers (effective bond length).}
Energies are measured in units of the reference 
energy scale $\epsilon\equiv 1$. Hence, the temperature $T$ scales in units 
of $\epsilon/k_\mathrm{B}$, where $k_\mathrm{B}$ is the Boltzmann 
constant (also set to unity in our simulations). 

Our model polymer with $N=55$ monomers is sufficiently long 
to identify the major structural effects without losing track of local and 
surface effects governing the different binding and folding mechanisms in the 
large parameter space covered by our structural analysis. The qualitative 
results do not depend significantly on the precise choice of the chain 
length. In this parameterization of the model, the ground-state structure of 
the free polymer is icosahedral~\cite{qlkpwb1}. However, for 
$\epsilon_\mathrm{S}\gg 0$ the interaction with the structurally incompatible 
hexagonal layer does not support quasicrystalline order in the polymer 
conformation. The formation of compact, adsorbed polymer conformations 
is governed by the recognition of the hexagonal structure of the 
 {monolayer} and 
adaptation to it.

The energy associated with the interaction of the 
polymer with the laterally infinitely extended hexagonal monolayer can 
be written as 
\begin{equation}
U_\mathrm{sub}(\textbf{X})=\epsilon_\mathrm{S}\sum_{i\ge 
2}^N 
U_\mathrm{S}(\mathbf{r}_i).
\end{equation}
The surface potential the $i$th 
monomer feels is given by 
\begin{equation}
U_\mathrm{S}(\mathbf{r}_i)=\sum_{a=1}^\infty 
U_\mathrm{NB}(r_{ia}),
\end{equation}
where $a$ is a formal index for the vertices in the 
hexagonal lattice and $r_{ia}$ is the distance between monomer $i$ and 
vertex $a$. The same cut-off condition is used as in the potential for 
non-bonded monomers.  {The nearest-neighbor distance between the 
attractive sites of the 
hexagonal layer is deliberately chosen so as to match the length scale $r_0$ 
of the effective 
bonds between monomers.}

\begin{figure}
\centerline{\includegraphics[width=8.8cm]{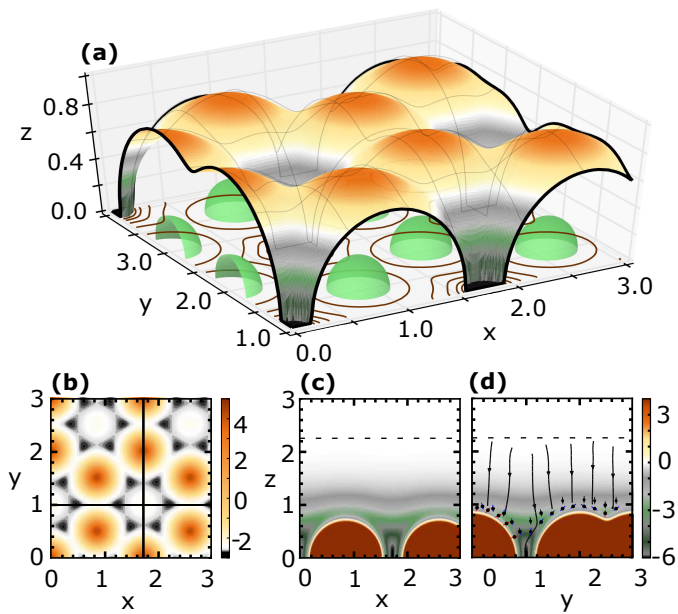}}
\caption{Equipotential surface $U_\mathrm{S}(\mathbf{r})\equiv 0$ (a) and 
three 
projections of the surface potential $U_\mathrm{S}(x,y,z)$ in the planes: (b) 
$z=0.81$, (c) $y=1$, and (d) $x=\sqrt{3}$. Note the saddle point at 
$y=2.5$ in (d) in addition to the stable potential 
minimum in the center of the hexagons.}
\label{fig:surfpot}
\end{figure}

The first monomer of the polymer chain is permanently 
attached to one vertex to prevent the polymer from escaping into the 
half-space above the substrate.
Figure~\ref{fig:surfpot} shows the equipotential surface 
$U_\mathrm{S}(\mathbf{r})\equiv 0$ as well as horizontal and vertical cross 
sections of the surface potential. The existence of the two minima is 
essential for the understanding of the 
structure formation mechanism of highly ordered and oriented adsorbed polymer 
conformations. 

Eventually, the total energy of a polymer conformation $\textbf{X}$ is given 
by
\begin{equation}
E(\textbf{X})=E_\mathrm{poly}(\mathbf{X})+E_\mathrm{sub}(\textbf{X}).
\end{equation}
Extensive parallel tempering replica-exchange~\cite{sw1,huku1,geyer1} 
simulations of this model were performed to generate high-quality statistics 
for thermodynamic quantities needed for the analysis of the transition 
behavior in a large space of adsorption parameter values and temperatures. 
Typical simulations consisted of up to 128 simulation threads non-uniformly 
distributed in temperature space and chosen as to  
guarantee sufficient overlap of energy histograms necessary for an adequate 
exchange probability between simulation threads at neighboring temperatures. 
For the interpolation of results at intermediate temperatures, the multiple 
histogram reweighting 
procedure~\cite{swendsen1,swendsen2} was employed.
\begin{figure}
\centerline{\includegraphics[width=8.8cm]{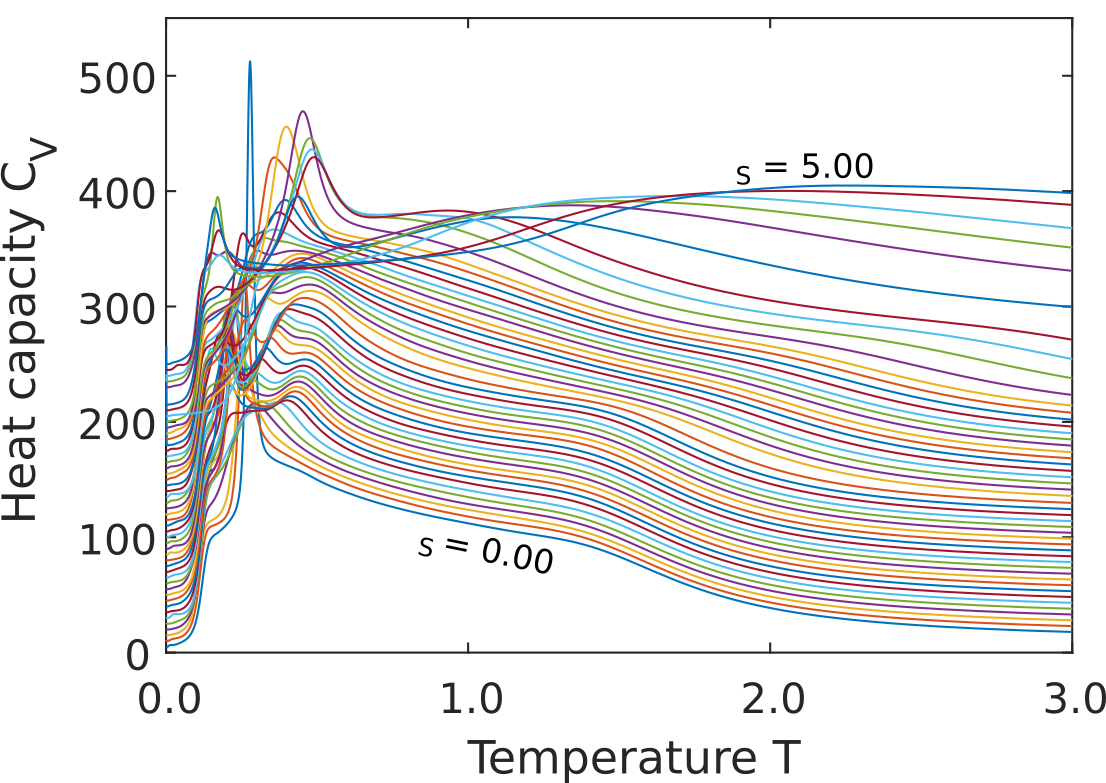}}
\caption{Illustrative example for the transition complexity as indicated by 
features 
like peaks and ``shoulders'' (concave regions) in the heat-capacity
curves $C_V(T)$ for an array of the adsorption strength parameter 
$\epsilon_\mathrm{S}$ in the interval $\epsilon_\mathrm{S}\in [0.0,5.0]$. The 
curves are shifted vertically.}
\label{fig:specheat}%
\end{figure}

\section{Statistical analysis} We analyzed a multitude of energetic and 
structural thermodynamic quantities 
and their fluctuations to identify pronounced features hinting to structural 
transitions in the system. The canonical statistical expectation value of a 
quantity $O(\textbf{X})$ is given by 
\begin{equation}
\langle 
O\rangle(T)=\frac{1}{Z_\mathrm{can}}\int {\cal 
D}\mathbf{X}\,O(\mathbf{X})\exp[-\beta E(\mathbf{X})],
\end{equation}
where
\begin{equation}
Z_\mathrm{can}=\int {\cal D}\mathbf{X}\exp[-\beta 
E(\mathbf{X})]
\end{equation}
is the canonical partition function at $\beta 
=1/k_\mathrm{B}T$ and ${\cal D}\mathbf{X}$ is the formal integral 
measure of all degrees of freedom in the space of polymer conformations 
$\mathbf{X}$. Thermal fluctuations are represented best by means of the 
response quantities (fluctuations)
\begin{equation}
\frac{d\langle O\rangle}{dT}=\frac{1}{k_\mathrm{B}T^2}\left(\langle 
OE\rangle-\langle O\rangle\langle E\rangle \right).
\end{equation}
Measurements of expectation values and fluctuations were performed in 
simulations for adsorption strength parameter values 
$\epsilon_\mathrm{S}\in[0.0,5.0]$. 

Figure~\ref{fig:specheat} shows the 
heat-capacity curves $C_V(T)=d\langle E\rangle/dT$ 
for all simulated model parameter values as an example for the challenging 
task to consistently locate potential transition points. Other quantities 
such as the radius of gyration and components of the gyration tensor, 
end-to-end distance, numbers of contacts between pairs of monomers and
monomers and lattice sites, and the center-of-mass location above the 
substrate have been used as more specific order parameters. Extremal 
fluctuations of these quantities help locate and characterize the
various transitions.

\section{Structural hyperphase diagram} 
\begin{figure}
\centerline{\includegraphics[width=8.8cm]{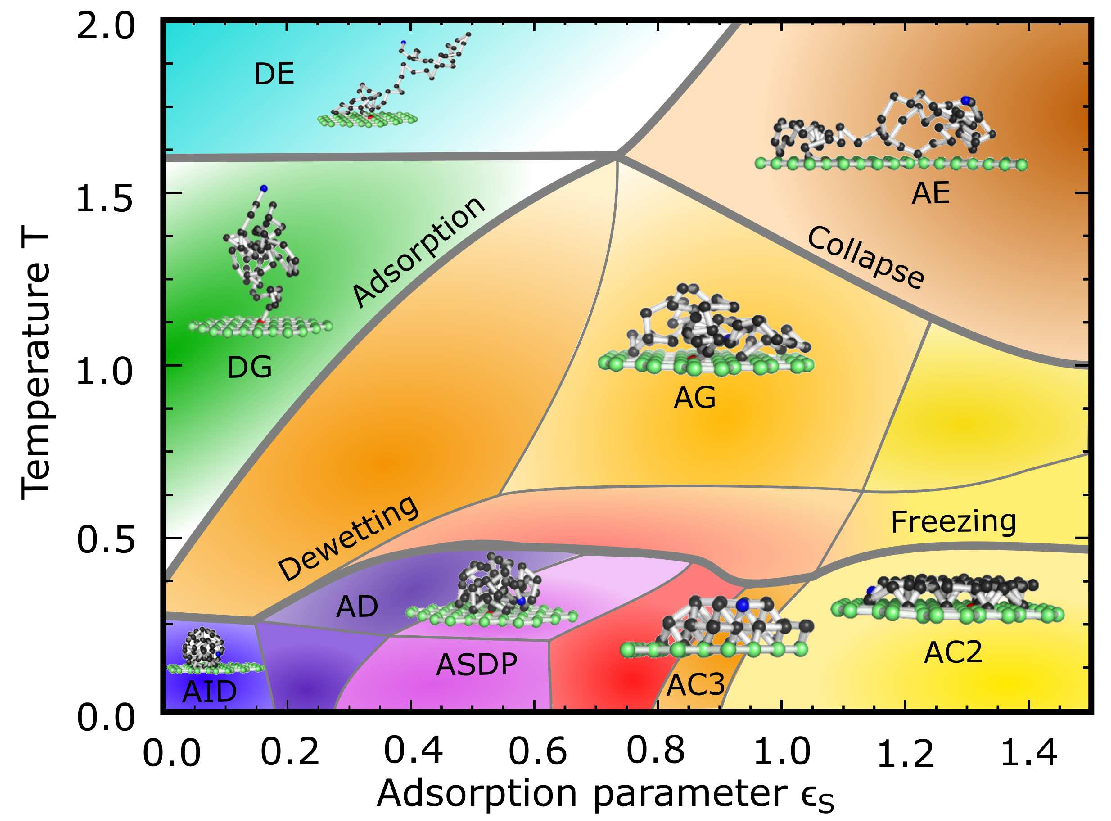}}
\caption{Hyperphase diagram of a flexible polymer  {with 55 monomers}
interacting with a 
hexagonal 
 {monolayer}. Representative polymer conformations are shown for 
major phases. See 
text for the description of the individual conformational 
phases.  {Since the studied system is finite, transition bands rather 
than lines indicate the
uncertainty in locating transition points when analyzing multiple response 
quantities by canonical statistical analysis. It should also be noted that 
whereas the precise location of transition lines depends on the polymer chain 
length, the general, qualitative structure of the phase diagram does not 
significantly change for larger systems.}}
\label{fig:pd}%
\end{figure}

The major result of the detailed 
statistical analysis is the 
$T$-$\epsilon_\mathrm{S}$ hyperphase diagram of the system, shown in 
Fig.~\ref{fig:pd} for $\epsilon_\mathrm{S}<1.5$. As expected, the polymer 
conformation is most compact at 
very low temperatures ($T<0.25$) and sufficiently small surface attraction 
($\epsilon_\mathrm{S}<0.2$). The structural phase of the polymer can be best 
characterized as ``adsorbed icosahedral droplet'' (AID). Even at 
$\epsilon_\mathrm{S}=0$ the polymer is forced to 
reside in close proximity to the substrate despite the lack of
attractive interaction. The tendency to maximize density while being grafted 
creates an effective spatial constraint. Increasing the temperature beyond 
the threshold $T\approx 0.38$, thermal fluctuations force the solid 
conformation to melt and an 
effective lift-off of the polymer is indicated by the structural order 
parameters we analyzed. In this phase polymer conformations are desorbed 
globules (DG). Upon further heating, entropic effects lead to the expected 
transition to desorbed expanded (DE) or random-coil structures. Adsorption of 
these polymer structures is enforced if the surface attraction of the 
substrate is large enough to overcompensate the effective (entropic) lifting 
force. 

Like in the desorbed 
phases, lowering the temperature leads to the collapse of adsorbed expanded 
(AE) polymer structures, and adsorbed globular (AG) conformations form. The 
AG phase is a complex phase, which is divided in subphases dominated by
globular droplets (below $\varepsilon_\mathrm{S}<0.6$), semispherical 
droplets ($0.6<\varepsilon_\mathrm{S}<1.2$) and strictly double-layered 
disordered film-like structures ($\varepsilon_\mathrm{S}>1.2$), which 
represent the thinnest possible structures for this system (fluctuations in 
the direction perpendicular to the substrate cease). Note the 
additional subphases around $T\approx 0.5$ just above the dewetting/freezing 
transition lines, which are not present in phase diagrams of adsorption for 
systems with uniform substrates~\cite{mbj1}. These additional subphases are 
caused by the adaptation of the flexible polymer to the 
hexagonal lattice structure of the  {monolayer}. The surface 
layer of the polymer 
conformation takes on an ordered, organized shape, whereas the upper 
layers are still globular.
Eventually, surface-layer structure formation is distinctly 
different at hexagonal sheets than near uniform substrates without 
inherent lattice structure~\cite{mbj1}. 
Before we discuss these features in more detail, which, in fact, render 
globular and frozen double-layer phases indistinguishable for adsorption 
strengths $\epsilon_\mathrm{S}>2.0$, we first focus on 
the conformational properties of compact polymer structures for 
$\epsilon_\mathrm{S}<1.5$.

Similar to the heat-capacity curves shown in Fig.~\ref{fig:specheat}, 
fluctuation properties of all structural parameters analyzed in this study 
exhibit enormous complexity in the low-temperature regime associated with the 
structurally most compact phases for 
$\epsilon_\mathrm{S}<1.5$. This is due to the competing structural 
organization schemes in the surface layer of the polymer conformation in 
direct contact with the hexagonal sheet and the structure 
formation processes in upper, three-dimensionally extended, sections of 
compact conformations. Whereas the AID phase is unaffected by the sheet 
pattern, the adjacent phases of adsorbed droplets (AD) and adsorbed 
semi-spherical droplets (ASD) are split into two almost horizontal subphases. 
Similarly to the globular phase, dominant structures in the subphases differ 
in the arrangement of monomers in the layer closest to the hexagonal sheet.

At temperatures $T>0.25$, the specific lattice structure of the sheet 
is not recognized and polymer 
crystallization proceeds as if the substrate was unstructured. However, at 
temperatures $T<0.25$, reordering the monomers 
in accordance with the hexagonal sheet structure has small energetic 
advantages over entropic freedom and the appearance of the solid structures 
slightly changes. 
An intermediate phase, located roughly in the interval 
$0.6<\epsilon_\mathrm{S}<0.8$ represents an effective wetting/dewetting 
transition between most compact droplet and the adsorbed compact layered, 
crystalline phases with multiple (AC3) and eventually two (AC2) layers. 
Layering and adopting crystalline features thus occur
once the surface adsorption strength is either competitive or dominates 
over inter-monomer attraction, i.e., for $\epsilon_\mathrm{S}>0.8$.

In these phases, 
monomers increasingly prefer contact with sheet vertices and 
intrude into the hexagonal sheet. However, since the hexagonal 
lattice is rigid, the distance between centers of adjacent hexagons is 
$\sqrt{3}r_0\approx 1.73r_0$ and the FENE potential prevents 
overstretching the bonds between bonded monomers beyond $r_0+R\approx 
1.43r_0$. Thus, at least three monomers are needed to fill centers of 
hexagons, with an intermediate monomer residing above the surface. The 
polymer must \emph{build bridges} to achieve optimal packing. This is made 
possible by the conditionally stable surface potential minima between vertices 
that are nearest neighbors in the hexagonal sheet [cf.\ 
Figs.~\ref{fig:surfpot}(a, d)]. This also means that, for surface 
attraction strengths $\epsilon_\mathrm{S}>1.0$, the most compact film-like 
polymer conformation must have two layers (AC2). This is 
substantially different from the adsorption behavior at unstructured 
substrates~\cite{mbj1}. Multiple-layer structures (in AC3) have a hexagonal 
closest packing (hcp) arrangement in striking contrast to polymer 
conformations at unstructured substrates which prefer forming face-centered 
cubic (fcc) structures. The hexagonal sheet truly imposes a structural 
constraint on the polymer, which leads to the formation of highly oriented, 
ordered conformations.
\begin{figure}
\centerline{\includegraphics[width=7cm]{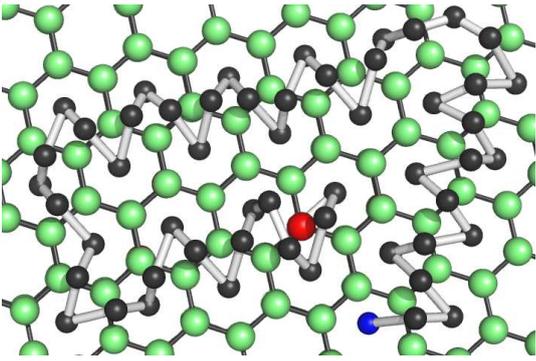}}
\caption{Prominent example of a polymer conformation in the AZ nanorail 
phase. Since 
bonded 
monomers cannot reside in neighboring cells, an intermediate monomer is 
required to build a bridge across the hexagonal vertex bonds, forming 
stable segments of straight ridge lines (``rails''). The grafted monomer is 
colored in 
red, the terminal monomer of the chain in blue.}
\label{fig:az}%
\end{figure}
\section{Nanorail phase} The most amazing of these structures are formed 
if the vertex-monomer 
attraction exceeds but not entirely eclipses monomer-monomer attraction. The 
result are polymer conformations organized in a bundled zig-zag pattern 
reminiscent of planar helix bundles, but folded upward as the polymer cannot 
penetrate the substrate. A representative example is shown in 
Fig.~\ref{fig:az}. This arrangement and ordering is energetically 
preferential, because monomers located in the centers of neighboring 
hexagonal cells feel sufficient attraction to form strands in close 
proximity to each other. 
The upper layer is composed of parallel or 
specifically angled linear ridges that form a distinct pattern by 
themselves. Other structures can also form, but all have the distinctive 
adsorbed ``zig-zag'' (AZ) bridge pattern in common. The zig-zag pattern 
of the strands is stabilized by the 
attractive interaction of the monomers located in the ridge.
 
The AZ phase dominates 
at low temperatures for surface attraction strengths 
$\epsilon_\mathrm{S}>2$. Upon increasing temperature  
the only weakly bound monomers located in hexagon centers are entropically 
dissolved first, but the zig-zag pattern remains intact even at significantly 
higher temperatures. For example, at $\epsilon_\mathrm{S}=3$, the transition 
from the highly oriented AZ conformations to the phase of adsorbed expanded 
zig-zag structures (AEZ) occurs at about $T\approx 1.5$. This means the 
hexagonal sheet is recognized by the polymer even at comparatively 
high temperatures until the polymer chain desorbs at about $T\approx 6.0$, 
where the majority of sheet-layer monomers are lifted off the hexagon 
centers for the gain of translational entropy at the expense of energetic 
benefits. It is worth noting that the ridges formed in 
the AZ phase can be straightened out by imposing energetic penalties 
for deviations from the optimal, alternating torsion angles 
dictated by the hexagonal lattice structure, $60^\circ$ and $120^\circ$. 
The dominant bond angles we found for the straight segments are approximately 
$50^\circ$ at the bridge foundation in the hexagon centers and about 
$100^\circ$ at the crest of the bridge. These values cannot be explained 
easily by geometric considerations, because the bonds between monomers have 
to overstretch by more than $10\%$ for large adsorption strengths. Under 
these conditions, the extremely stable ridge line is 
exactly linear and forms a ``rail'' along which molecular transport 
is possible.

\section{Summary and conclusions} Our study of the adsorption behavior of a 
flexible
polymer at a hexagonally structured monolayer fixed 
on a solid substrate revealed a multitude of unique structural phases. 
Structure recognition and adaptation are  
mechanisms that lead to the formation of ordered and 
stable polymer conformations, which differ from adsorbed polymer structures 
at uniform substrates. A self-organized bridge-building mechanism 
enables the optimal 
coating of the hexagonal sheet and the formation of highly organized strands 
of monomers in parallel and angled arrangements. In actual applications, 
these polymer coats may help stabilize hexagonal sheets and 
nanotubes and add specific function. The high orientation of the polymer 
strands with their distinctive rail-like ridges~\cite{nanorails1,nanorails2} 
enables directed molecular or 
vesicular transport on hexagonal substrates. Our results also motivate 
promising future studies such as nanoscale threading through penetrable 
sheets. 

\acknowledgments
B.L.\ acknowledges support by the Studienstiftung des deutschen Volkes 
(German Academic Scholarship Foundation). 
This work has been partially supported by the NSF under Grant No.\ 
DMR-1463241.

\end{document}